\documentclass[runningheads]{llncs}

\usepackage{eccv}

\usepackage{eccvabbrv}

\usepackage{graphicx}
\usepackage{booktabs}
\usepackage{booktabs}
\usepackage{multirow}
\usepackage{makecell}
\usepackage{graphicx}
\usepackage[table]{xcolor}
\usepackage{booktabs}
\usepackage{multirow}
\usepackage{makecell}
\usepackage{graphicx}
\usepackage{xcolor}
\usepackage{booktabs, multirow, makecell, graphicx}

\usepackage[accsupp]{axessibility}  % Improves PDF readability for those with disabilities.
\usepackage{xurl}

\usepackage{hyperref}

\usepackage{orcidlink}

\definecolor{mj}{rgb}{0.21,0.49,0.74}
\begin{document}

\title{MoZoo:Unleashing Video Diffusion power in animal fur and muscle simulation} 

\titlerunning{Abbreviated paper title}

\author{Dongxia Liu\inst{1,4} \and Jie Ma\inst{4} \and Xiaochen Yang\inst{2} \and Jiancheng Zhang\inst{4} \and Bin Xia\inst{3} \and Zhehan Kan\inst{1} \and Nisha Huang\inst{1} \and Jun Liang\inst{4} \and Wenming Yang\inst{1}$^\star$ \and Jing Li\inst{4}}

\authorrunning{D. Liu et al.}
% If there are more than two authors, 'et al.' is usessd.

\institute{
\begin{tabular}{c}
$^1$ Tsinghua University \quad $^2$ University of Glasgow \quad $^3$ The Chinese University of Hong Kong \\
$^4$ HUIJING Digital Media \& Entertainment Group \\
\vspace{2mm}
\centerline{ \textbf{Project Page: \url{https://dongxialiu15.github.io/MoZoo/}}}
\end{tabular}
}

\maketitle

\vspace{-3ex}
\begin{figure}[h] % [t] 表示建议放在页面顶部，也可以用 [h] 表示尽量放在当前位置
    \centering
    \includegraphics[width=\linewidth]{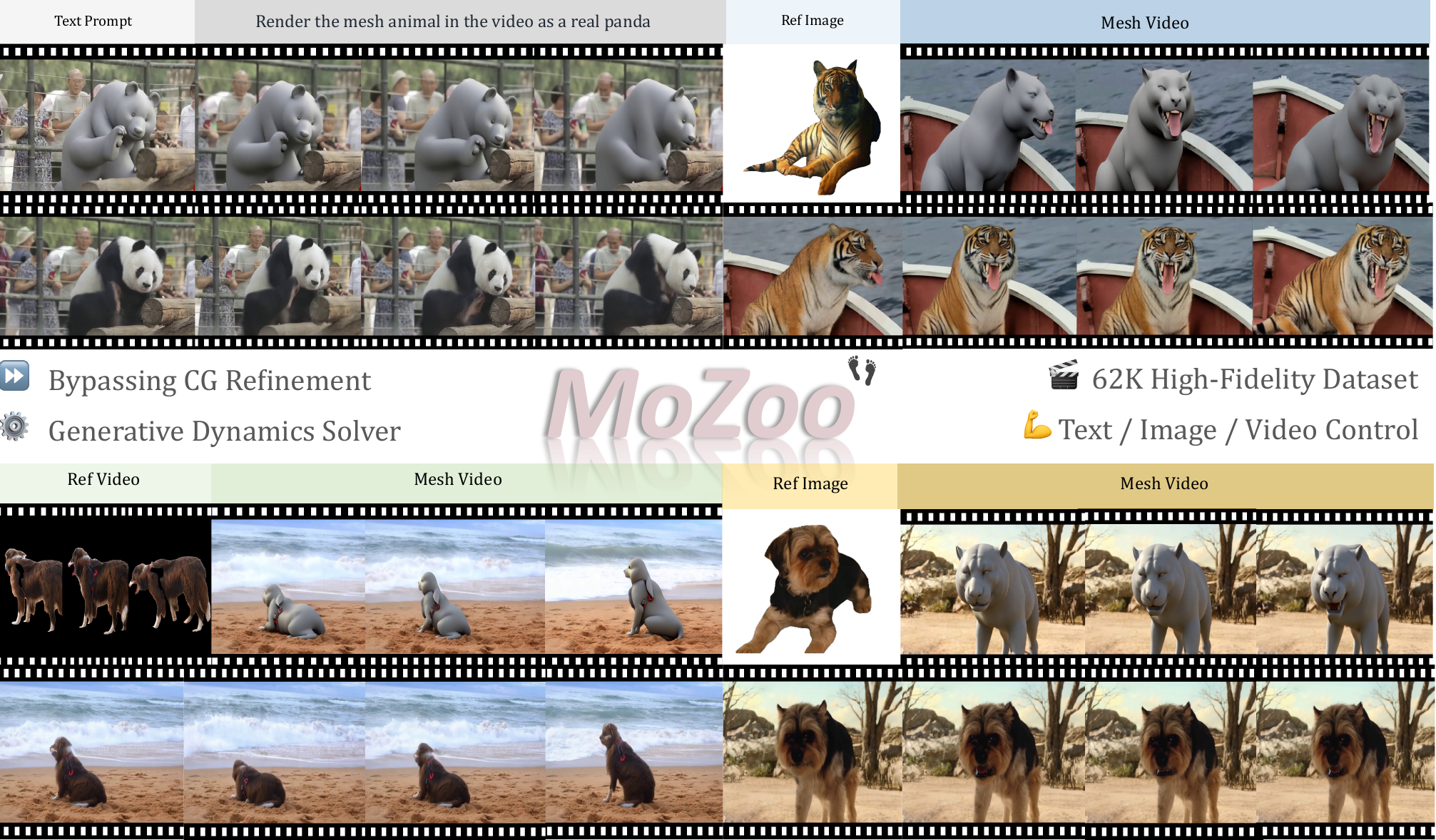} 
    \caption{\textbf{Visualization for our MoZoo results.} Given a source mesh video, MoZoo synthesizes photorealistic animal videos by transferring textures from multimodal references—including text prompts, images, and videos. The framework maintains precise motion synchronization and structural consistency even when transferring attributes across different species.}
    \label{fig:teaser} % 这里的标签用于正文引用
\end{figure}

\vspace{-3ex}
\begin{abstract}
The creation of cinematic-quality animal effects necessitates the precise modeling of muscle and fur dynamics, a process that remains both labor-intensive and computationally expensive within traditional production workflows. While generative diffusion models have shown promise in diverse artistic workflows, their capacity for high-fidelity animal simulation remains largely unexploited. We present \textbf{MoZoo}, a generative dynamics solver that bypasses conventional refinement to synthesize high-fidelity animal videos from coarse meshes under multimodal guidance. We propose Role-Aware RoPE (RAR-RoPE) which employs role-based index remapping to synchronize motion alignment while decoupling reference information via fixed temporal offsets. Complementing this, Asymmetric Decoupled Attention partitions the latent sequence to enforce a unidirectional information flow, effectively preventing feature interference and improving computational efficiency. To address the scarcity of high-quality training data, we introduce \textbf{MoZoo-Data}, a synthetic-to-real pipeline that leverages a rendering engine and an inverse mapping approach to construct a large-scale dataset of paired sequences. Furthermore, we establish \textbf{MoZooBench}, a comprehensive benchmark with 120 mesh-video pairs. Experimental results demonstrate that MoZoo achieves high-fidelity fur simulation across diverse animal skeletons and layouts, preserving superior temporal and structural consistency.
\end{abstract}

% At the core of our framework is an asynchronous dual-feature exchange attention mechanism, which explicitly decouples structural guidance from textural information. This design allows the model to faithfully follow the target motion while capturing the intricate, volumetric details of the reference fur.

% \include{main/intro}
\section{Introduction}
\label{sec:intro}
\begin{figure}[h] % [t] 表示建议放在页面顶部，也可以用 [h] 表示尽量放在当前位置
    \centering
    \includegraphics[width=\linewidth]{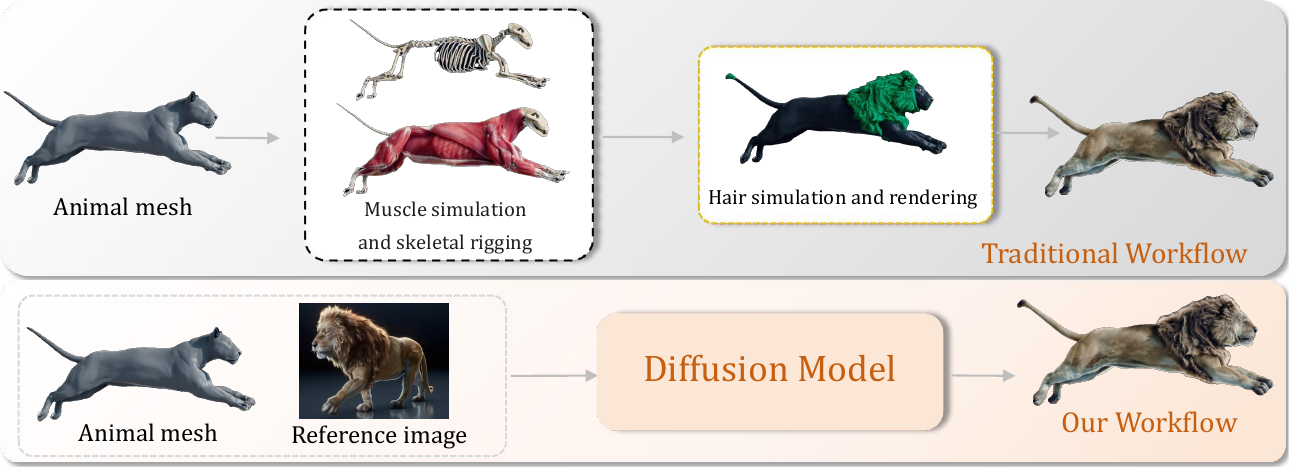} 
    \caption{Comparison of production workflows. Traditional pipelines (top) require a sequence of complex simulation stages, including muscle rigging and hair rendering. In contrast, MoZoo (bottom) streamlines this into a single generative dynamics solving process, synthesized directly from a mesh and a reference image.}
    \label{fig:First} % 这里的标签用于正文引用
\end{figure}

Creating lifelike animal effects is essential for  cinematic realism and immersive virtual experiences, which necessitates the precise modeling of anatomy, muscle deformation, and fur dynamics~\cite{10.1145/2775280.2792559}.s shown in Fig.~\ref{fig:First}, the traditional production pipeline relies on a sequence of labor-intensive simulation stages, primarily encompassing skeletal rigging, muscle modeling, and subsequent high-fidelity fur synthesis. Most development effort is concentrated on these physically-based refinement steps. TThis process requires experienced artists to build complex systems for skeletal motion and fur interactions, often necessitating manual adjustments of subtle deformations to ensure cinematic quality. The resulting technical complexity and resource demands create a significant barrier for many creators.
% In film and visual effects production, high-fidelity rendering of animal fur is essential for achieving cinematic realism and immersive virtual experiences. Conventional workflows rely on meticulous manual grooming and frame-by-frame adjustments, demanding immense human expertise and vast  resources. These high costs pose significant challenges, limiting access for many potential creators.

Recent significant breakthroughs in generative AI have facilitated the integration of diffusion models \cite{wan2025, HaCohen2024LTXVideo, yang2024cogvideox, li2025tooncomposerstreamliningcartoonproduction,guo2026dreamidvbridgingimagetovideogaphighfidelity} into diverse artistic workflows. However, as existing methods\cite{li2025adaviewplanneradaptingvideodiffusion,liu2026animatescenecameracontrollableanimationscene,shi2025drivemesh4dlatent} are primarily optimized for human subjects and general consumer applications, the accurate simulation of  fur dynamics remains largely unexplored. The core reason lies in the lack of sufficient high-quality paired data, and the generated videos often have noticeable synthetic artifacts, lacking realism and fine details. Current reference-based video methods\cite{huang2025refaccadeeditingobjectgiven,xue2025standinlightweightplugandplayidentity,guo2026wildactorunconstrainedidentitypreservingvideo} often rely on a single reference image to guide content generation. However, the geometric complexity of fur, characterized by dense fiber arrangements and intricate light-hair interactions, is difficult to capture from a single 2D view~\cite{chang2026hairweaverfewshotphotorealistichair,sklyarova2026neuralfuranimalfurreconstruction,xia2026omegaavataroneshotmodeling360deg}. Compared to static images or textual descriptions, reference videos provide dynamic textural nuances and enhanced spatial consistency.

% Notably, while these techniques excel in surface textures, the high-fidelity generation of animal fur has not been effectively explored. The core reason lies in the lack of high-quality paired data, and the generated videos often have noticeable synthetic artifacts, lacking realism and fine details. Existing methods \cite{huang2025refaccadeeditingobjectgiven} rely mainly on pixel-level cues from a single reference image. However, the geometric complexity of fur, characterized by dense fiber arrangements and intricate light-hair interactions, is difficult to capture from a single 2D view. Compared to static images or textual descriptions, reference videos provide dynamic textural nuances and spatial consistency. Building on these observations, we introduce \textbf{Mozoo}, a framework for high-fidelity animal fur synthesis. Given a mesh video and an arbitrary animal reference video, our method transfers textures from the reference to the mesh sequence. By utilizing multi-frame information from the reference video, the framework maintains both motion dynamics and fine appearance details throughout the generation process.

To address existing data bottlenecks, we introduce \textbf{Mozoo-Data}, a comprehensive large-scale dataset constructed through a specialized generative pipeline. The process begins with a diverse collection of animal meshes and animations from public sources, which are rendered in Unreal Engine 5 (UE5) to produce paired sequences of coarse meshes and RGB videos. However, these synthetic RGB videos often lack the intricate biological nuances of muscle deformation and fur dynamics observed in real animals, making them insufficient for direct training. We address this limitation by training an inverse generative model on the synthetic data to map RGB sequences back to their corresponding coarse mesh representations. This model is subsequently applied to high-definition wildlife documentaries and other real-world footage to extract estimated precise coarse mesh sequences. The resulting dataset consists of high-fidelity real videos paired with their algorithmically derived mesh counterparts.

% we first designed a Render data pipeline to generate customized data based on a rendering engine. Then, to tackle the lack of real animal data, we trained a mesh generation model guided by the first frame using the rendered data. Finally, we collected videos of real animals and used Flux2kelin\cite{flux-2-2025} to edit the subjects in the first frame into mesh structures, generating paired data with our trained mesh generation model.

Building upon our data pipeline, we develop \textbf{Mozoo}, a generative dynamics solver designed to bypass the refinement stage of traditional computer graphics~(CG) pipelines. MoZoo accepts a coarse mesh video and is conditioned on multi-modal guidance, such as text prompts, reference animal images, or videos. To address the varying alignment requirements of different conditions, we introduce ~\textbf{Role-Aware RoPE (RAR-RoPE)}. This method synchronizes temporal indices for the target and mesh videos while decoupling reference content through a fixed temporal offset, thereby resolving the artificial temporal priors between the reference and mesh sequences. We further propose~\textbf{Asymmetric Decoupled Attention}, which partitions the latent sequence into four functional segments. By constraining each target query to its frame-synced mesh counterpart and isolating reference branches from the noisy target sequence, this mechanism ensures the explicit separation of structural guidance from textural information. This approach enables high-fidelity feature integration while significantly reducing the computational overhead compared to standard full-attention models.

% The core of our framework is an asynchronous dual-feature exchange attention mechanism. Unlike previous methods that treat structure and texture as a unified condition, the asynchronous dual-feature exchange attention explicitly decouples the structure guidance provided by the mesh video from the texture information of the reference video. By injecting this asymmetric information into the self-attention layers of the diffusion model, our method ensures that the generated video strictly follows the mesh motion poses while capturing the texture features of the reference animal video. At the same time, by applying temporal shifts, we break the pseudo-prior constraints between the reference video and the mesh video, significantly enhancing the flexibility of texture mapping and the stability of motion.

Through the aforementioned improvements, our method can effectively generate fine animal fur. To validate the effectiveness of Mozoo, we introduce a comprehensive benchmark \textbf{MozooBench}, which consists of 120 mesh videos and comes with paired real animal and corresponding reference animal videos. Experiments demonstrate our model's high-efficiency texture transfer capability. As shown in Fig.~\ref{fig:teaser}, MoZoo can generate detailed textures based on reference animal videos, even on subjects of different species, while maintaining structural consistency, showcasing its robust generalization capabilities across diverse biological morphologies and motion patterns.

In summary, our main contributions are as follows:
\begin{itemize}
\item We propose \textbf{Mozoo}, a novel video diffusion framework that replaces the most computationally expensive simulation stages in traditional CG pipelines, directly generating high-fidelity animal videos with realistic muscle and fur dynamics from coarse mesh inputs. Mozoo supports multimodal guidance, enabling flexible control over video generation through various reference sources.
\item We introduce \textbf{Mozoo-Data}, a large-scale dataset that includes synthetic UE assets with real-world animal footage. By employing an inverse mapping approach, we extract coarse mesh sequences from real-world videos to capture high-fidelity biological dynamics. This dataset, together with our benchmark \textbf{MozooBench}, will be made publicly available.
\item Extensive quantitative and qualitative experiments demonstrate that MoZoo significantly outperforms state-of-the-art methods in terms of photorealism, temporal consistency, and multi-modal synthesis flexibility.
\end{itemize}

\section{Related Work}

\subsection{Video generation models}
The rapid development of diffusion models~\cite{ho2020denoisingdiffusionprobabilisticmodels,song2022denoisingdiffusionimplicitmodels,dhariwal2021diffusionmodelsbeatgans} has significantly advanced research into large-scale video foundation models~\cite{HaCohen2024LTXVideo,wan2025,yang2024cogvideox,zheng2024opensorademocratizingefficientvideo,kong2024hunyuanvideo}. While early work~\cite{xing2023makeyourvideocustomizedvideogeneration,singer2022makeavideotexttovideogenerationtextvideo,ho2022videodiffusionmodels} primarily focused on extending text-to-image architectures to ensure temporal coherence, reference-based generation~\cite{zhang2023addingconditionalcontroltexttoimage,ye2023ipadaptertextcompatibleimage} has emerged as a pivotal research direction for improving controllability. Current reference-guided methods~\cite{guo2026dreamidvbridgingimagetovideogaphighfidelity,he2024idanimatorzeroshotidentitypreservinghuman,yuan2025identitypreservingtexttovideogenerationfrequency,xue2025standinlightweightplugandplayidentity} mostly rely on a single reference image to guide the generative process. For example, VideoCrafter~\cite{chen2024videocrafter2} and DynamiCrafter~\cite{xing2023makeyourvideocustomizedvideogeneration} use pre-trained encoders to extract semantic guidance from a single reference frame. Following the growth of In-Context Learning~\cite{ju2025fullditmultitaskvideogenerative,tan2025ominicontrolminimaluniversalcontrol}, recent studies have begun concatenating reference features along the temporal axis to enable conditional generation. However, these prior approaches often overlook the intrinsic multi-frame and multi-view cues naturally available within diverse video sources. Our proposed method systematically exploits these latent cues to achieve a more comprehensive and thorough extraction of reference features.

\subsection{Hair and muscles simulation and rendering}
Achieving high fidelity hair and muscles simulation and rendering remains a formidable challenge in computer graphics. While professional pipelines utilize specialized  suites~\cite{houdini}, intricate shading models~\cite{10.1145/2775280.2792559}, and physically-based path tracing~\cite{lin2025controlhairphysicallybasedvideodiffusion} to deliver compelling results, the sheer geometric complexity and strand count render manual modeling workflows exceptionally time-consuming. Despite significant strides in physics-based~\cite{10.1145/882262.882345,10.1145/1399504.1360630,10.1145/1360612.1360631} and data-driven approaches~\cite{wang2023neuwigsneuraldynamicmodel,luo2024gaussianhairhairmodelingrendering}, an optimal trade-off between reconstruction automation and visual fidelity remains elusive. Recently, the advent of learned priors~\cite{rosu2022neuralstrandslearninghair,sklyarova2023neuralhaircutpriorguidedstrandbased,wu2024monohairhighfidelityhairmodeling} and anisotropic 3D Gaussians~\cite{zakharov2024humanhairreconstructionstrandaligned,Zhou_2024} has fundamentally enhanced hair representation. Concurrently, frameworks like ControlHair~\cite{lin2025controlhairphysicallybasedvideodiffusion} have demonstrated the potential of fine-tuned video diffusion models for directable hair motion. Recognizing that hair and muscles faces simulation and rendering structural challenges similar to hair reconstruction, this work proposes to integrate video diffusion priors with streamlined geometric representations. Our objective is to bridge the gap between low-cost production and high-end visual fidelity, ultimately matching or exceeding the quality of state-of-the-art industrial pipelines.

% \begin{figure}[h] % [h] 表示建议放在页面顶部，也可以用 [h] 表示尽量放在当前位置
%     \centering
%     \includegraphics[width=\linewidth]{paper-template-Latest/figure/datapipeline.pdf} 
%     \caption{Overview of datapipeline}
%     \label{fig:datapipeline} % 这里的标签用于正文引用
% \end{figure}

\section{Preliminary}
Video diffusion model~\cite{wan2025} uses the Rectified Flow~\cite{liu2022flowstraightfastlearning} framework to model the generation process and employs a transformer~\cite{peebles2023scalablediffusionmodelstransformers} as the denoising network. The input video $V$ will first be mapped into a latent representation $z_0$ using a video VAE encoder~\cite{kingma2022autoencodingvariationalbayes}. Rectified Flow is then used to perform linear interpolation between noise and data in the forward process. At any timestep $t \in [0, 1]$, the noisy latent $z_t$ is defined as:
\begin{equation}
    z_t = (1 - t)z_0 + t\epsilon,
\end{equation}

During the training phase, the Diffusion Transformer~(DiT) model $v_\theta$ is trained to directly regress the target velocity $u_t = \epsilon - z_0$, which represents the direction from data to noise. The training objective $\mathcal{L}_{FM}$ is formulated as:
\begin{equation}
    \mathcal{L}_{FM} = \mathbb{E}_{t, z_0, \epsilon, c} \| v_\theta(z_t, t, c) - u_t \|_2^2,
\end{equation}
where $c$ denotes the conditioning information such as text prompts or reference images.

In the inference phase, starting from $z_1 \sim \mathcal{N}(0, \mathbf{I})$, the latent $z_0$ is generated by solving the ODE $dz_t = v_\theta(z_t, t, c)dt$ from $t=1$ to $t=0$:
\begin{equation}
    z_{t-\Delta t} = z_t - v_\theta(z_t, t, c) \Delta t.
\end{equation}
The final video is reconstructed via a VAE decoder.
\section{Method}
\label{sec:Method}
Given a reference animal video and a source video featuring an untextured mesh proxy within a realistic scene, our goal is to render a lifelike subject by transferring the intricate textures from the reference video. To achieve this, we introduce \textbf{Mozoo}, a framework built upon the DiT-based video diffusion model, Wan2.1~\cite{wan2025}. In the following sections, we first introduce a new data pipeline to construct paired data (Sec.~\ref{sec:Mozoopipe}). Following the data construction, we detail our architectural enhancements to the DiT backbone. Specifically, we propose Role-Aware RoPE (RAR-RoPE) to resolve the alignment challenges inherent in multi-modal video sequences (Sec.~\ref{sec:bias}). Furthermore, we introduce Asymmetric Decoupled Attention, which partitions the latent sequence into specialized functional segments for high-fidelity feature integration (Sec.~\ref{sec:Attention}).

\begin{figure}[t] % [h] 表示建议放在页面顶部，也可以用 [h] 表示尽量放在当前位置
    \centering
    \includegraphics[width=\linewidth]{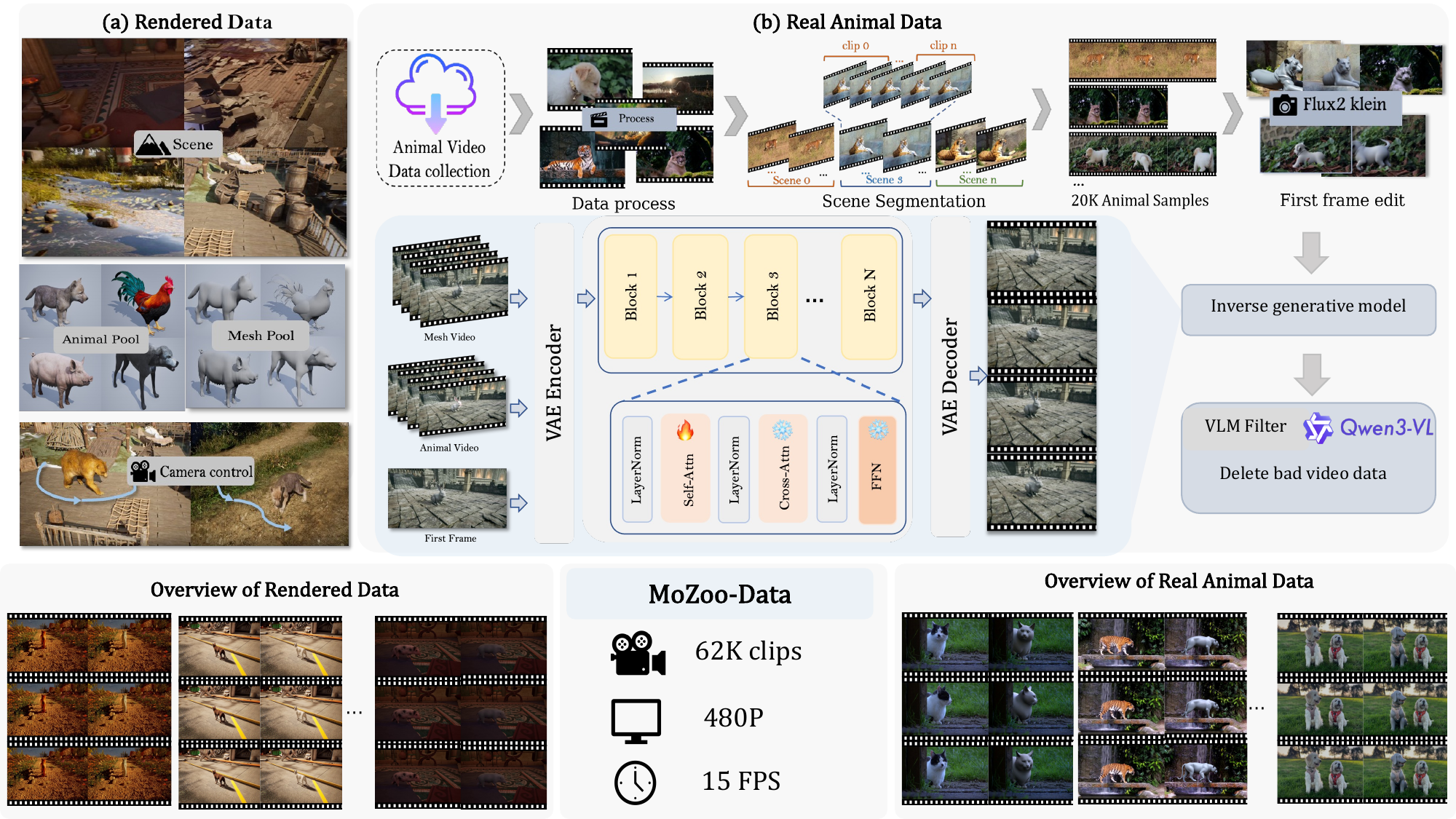} 
    \caption{Pipeline for Constructing MoZoo-Data from Synthetic and Real-World Sources. (a) Synthetic data generation utilizes Unreal Engine 5 with diverse animal assets, scenes, and camera trajectories. (b) Real-world footage is processed through scene segmentation and first-frame editing, followed by mesh extraction via an inverse generative model and quality filtering with a Vision-Language Model. The resulting dataset comprises 62K clips at 480P resolution and 15 FPS.}
    \label{fig:datapipeline} % 这里的标签用于正文引用
\end{figure}
\subsection{Data Pipeline}
\label{sec:Mozoopipe}
Training our task requires a video triplet comprising  a texture reference video $V_{\text{ref}}$, a mesh reference video $V_{\text{mesh}}$, and a target video $V_{\text{tar}}$. Within this framework, $V_{\text{tar}}$ and $V_{\text{mesh}}$ share identical motion trajectories, facial expressions, and background dynamics, while $V_{\text{ref}}$ serves as the primary reference for the fur characteristics of the target animal. 
Due to the commercial sensitivity of professional animation assets and the prohibitive costs of manual annotation, acquiring such paired data in the real world is both technically demanding and expensive. To address these challenges, we design a data pipeline that integrates virtual engine assets with real animal data, as illustrated in Fig.~\ref{fig:datapipeline}.

\subsubsection{Rendered Data}
We develop an automated rendering pipeline using the Unreal Engine 5 (UE5)~\cite{unrealengine55}. The pipeline incorporates various high-fidelity assets, including 3D scenes and mesh structures with their corresponding animal texture. By using rigged assets and predefined motion sequences, the system generates realistic behavioral interactions. To construct a high-fidelity dataset, we curate a diverse mesh pool consisting of fully-rigged animal geometries across various species. For each rendering instance, we sample a specific asset from the pool and place it within a complex 3D scene. To obtain the structural guidance $V_{\text{mesh}}$, we first render the sampled asset using a simplified, textureless shader to capture the raw geometric deformation. Subsequently, the corresponding target video $V$ is generated by applying photorealistic materials and fur textures from our animal pool onto the identical animated mesh structure. By maintaining synchronized skeletal motion and randomized camera trajectories across both rendering passes, we ensure that $V_{\text{mesh}}$ and $V$ are bit-perfectly aligned, providing a consistent geometric-to-visual mapping for training. To facilitate localized texture learning, we concurrently export per-frame binary segmentation masks $M$ via the engine’s custom stencil pass, providing precise pixel-wise annotation of the animal subject. For the reference source $V_{\text{ref}}$, we implement a cross-sequence sampling strategy to enforce the disentanglement of identity and motion. We extract the animal subject from a non-overlapping temporal segment of the same species and environment, but driven by a disjoint motion trajectory. 
\subsubsection{Real Animal Data}
However, existing rendering datasets are insufficient to meet the requirements of our task because the significant domain gap between synthetic and real-world videos often leads to degraded generation quality. To bridge this gap, we propose a video editing pipeline that transforms real-world animal video into  mesh video. We first train a mesh generation model guided by the first frame using available synthetic data. To construct the real-world animal dataset, we collect videos from Pexels~\cite{pexels} and perform standardized preprocessing, which includes renaming, resizing, and the removal of vertical videos to maintain a uniform resolution of $1280 \times 720$. Scene detection tools are used to identify and remove shot transitions, ensuring that each segment is semantically continuous and free of abrupt scene changes. These segments are then cropped to a fixed length of 81 frames. Building on this, we utilize Flux2 Kelin~\cite{flux-2-2025} to perform hair-removal editing on the subject in the first frame, converting it into a mesh structure. This edited frame and the source video are then provided as input to the mesh generation model to produce paired data. Finally, we employ SAM 3~\cite{carion2025sam3segmentconcepts} to extract instance masks and construct reference videos $V_{\text{ref}}$ by sampling subjects from different action clips of the same species within the same scene category.

\begin{figure}[t] % [t] 表示建议放在页面顶部，也可以用 [h] 表示尽量放在当前位置
    \centering
    \includegraphics[width=\linewidth]{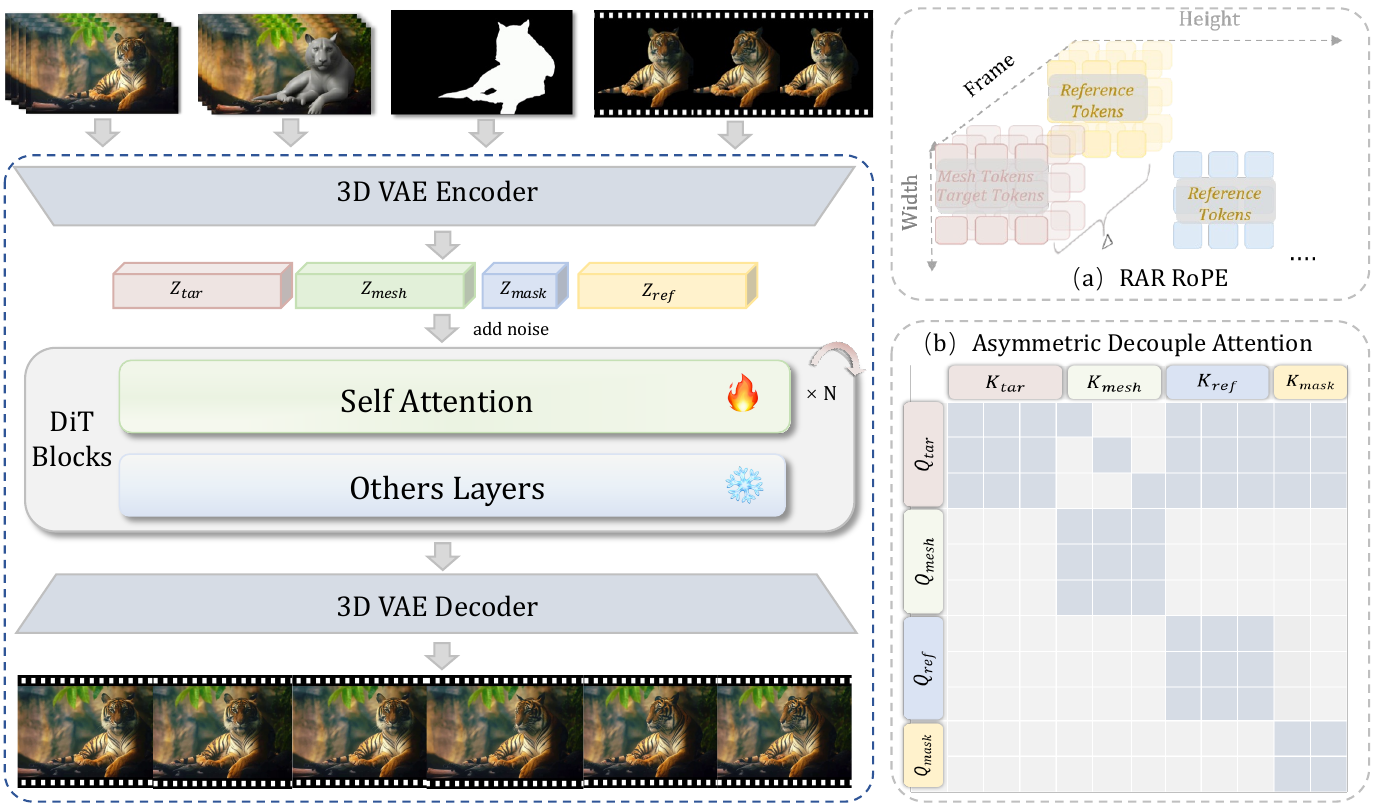} 
    \caption{Pipeline of the MoZoo Framework. (a) RAR Rope for target, mesh, and reference tokens across frame, width, and height dimensions. (b) The restricted attention matrix regulates information exchange among different latent components to maintain structural and appearance fidelity during the generation process.}
    \label{fig:pipeline} % 这里的标签用于正文引用
\end{figure}

\subsection{Role-Aware RoPE}
\label{sec:bias}
Our model requires four inputs: a mesh reference video $V_{\text{mesh}} \in \mathbb{R}^{n \times h \times w \times c}$ as source video, an animal reference video $V_{\text{ref}} \in \mathbb{R}^{n \times h \times w \times c}$ providing the reference fur textures, a first-frame mask of the mesh video $M_{\text{s}} \in \mathbb{R}^{1 \times h \times w \times 1}$, and a noisy target video $V_{\text{tar}} \in \mathbb{R}^{n \times h \times w \times c}$ during training (or Gaussian noise during inference). These inputs are encoded into latent representations $z_{\text{mesh}}, z_{\text{ref}},z_{\text{tar}} \in \mathbb{R}^{n' \times h' \times w' \times d}$ via a VAE, where $n$ and $h \times w$ represent the original temporal and spatial sizes, and $n', h', w'$ denote the latent dimensions. A straightforward baseline is to follow an in-context learning method~\cite{ju2025fullditmultitaskvideogenerative} after patching the latent representations, where all conditional tokens are concatenated along the temporal axis to form a unified sequence $\mathbf{x} = [\mathbf{x}_{\text{tar}}, \mathbf{x}_{\text{mesh}}, \mathbf{x}_{\text{mask}}, \mathbf{x}_{\text{ref}}]$, which is then processed by the full self-attention of the DiT backbone.

However, the standard 3D-RoPE assigns continuous indices to each condition along the temporal dimension, imposing a spurious temporal prior and  failing to account for the distinct alignment requirements of different conditions. Inspired by the observations of RoPE in in-context image generation~\cite{tan2025ominicontrolminimaluniversalcontrol}, we recognize that different input streams in video tasks exhibit heterogeneous alignment requirements. To address this,  we propose \textbf{Role-Aware RoPE (RAR-RoPE)} to assign indices based on token functions rather than linear order. Specifically, we synchronize temporal indices for the target and source videos to ensure precise motion alignment. To decouple reference information, we shift the temporal indices of reference prompts by a fixed offset $\Delta$, effectively positioning them prior to the noisy video tokens. Furthermore, spatial offsets in height and width are leveraged to distinguish individual reference  tokens, while the mask is anchored at index zero to maintain its structural role.
For a token $i$ with role $r$, its coordinate mapping $\mathcal{M}(i, r)$ is defined as:
\begin{equation}
\mathcal{M}(i, r) = 
\begin{cases} 
(t_i, h_i, w_i), & \text{if } r \in \{\text{tgt, src}\} \\
(t_i - \Delta, h_i, w_i), & \text{if } r = \text{ref (video)} \\
(-1, h_i, w_i), & \text{if } r = \text{ref (image)} \\
(0,  h_i, w_i), & \text{if } r = \text{msk}
\end{cases}
\label{eq:rar_rope}
\end{equation}

% A straightforward baseline is to share spatio-temporal embeddings across all video streams, following the principles of in-context structure-controlled generation, while fixing the temporal index of the mask to zero.
% Although the overall design is capable of achieving high-quality generation, this strategy is often limited by the rigid frame-to-frame correspondence imposed by shared temporal indexing. If the motion trajectories or expressions of $V_{\text{ref}}$ and $V_{\text{mesh}}$ are misaligned, the model fails to establish accurate semantic correspondences, leading to texture mapping failures and motion leakage. To eliminate this spurious prior, we shift the reference texture video by a temporal offset $\Delta$. 

\subsection{Asymmetric Decoupled Attention}
\label{sec:Attention}
While RAR-RoPE handles positional priors, vanilla self-attention still allows unrestricted bidirectional interaction, often leading to feature interference between mesh structural guidance and references fur.  The dominant edge gradients and low-frequency structural signals from the target mesh tend to overwhelm the subtle, high-frequency fur details in the reference latent. Additionally, concatenating several sequences significantly increases the number of tokens, resulting in a computational burden that is often impractical for real-world deployment.
% Standard self-attention allows for global interaction among all tokens. While this theoretically enables target video tokens to integrate mesh guidance and identity representations, unrestricted bidirectional attention often leads to interference between geometric and appearance features, which can degrade the output quality. Additionally, concatenating several reference sequences significantly increases the number of tokens, resulting in a computational burden that is often impractical for real-world deployment.

We propose \textbf{Asymmetric Decoupled Attention (ADA)} to address the limitations of vanilla self-attention in multi-source video generation. The input latent sequence $\mathbf{x}$ is partitioned into four segments: the target generation sequence $x_{\text{tgt}}$, mesh structural guidance $x_{\text{mesh}}$, global semantic mask $x_{\text{msk}}$, and animal reference $x_{\text{ref}}$. As shown in Fig.~\ref{fig:pipeline}, we implement an asymmetric masking strategy that allows the target queries $q_{\text{tgt}}$ to aggregate information from all modalities while keeping the guidance branches isolated. For each frame $i$ in the target sequence, the attention search space is specifically constrained:
\begin{equation}
K_{\text{comp},i} = [K_{\text{tgt}}, k_{\text{mesh},i}, K_{\text{msk}}, K_{\text{ref}}]
\end{equation}
In this formulation, $k_{\text{mesh},i}$ ensures that target frame $i$ attends exclusively to its corresponding structural key rather than the entire mesh sequence. This frame-wise alignment maintains strict temporal synchronization, while $K_{\text{ref}}$ and $K_{\text{msk}}$ are processed globally to provide consistent style and regional constraints.

To prevent the noise in the target sequence from corrupting the clean guidance signals, the reference and structural branches perform feature extraction exclusively within their own subspaces. The outputs for the reference and structure are computed as:
\begin{equation}
x_{\text{ref}} = \text{Attn}(q_{\text{ref}}, k_{\text{ref}}, v_{\text{ref}}) \quad \text{and} \quad x_{\text{str}} = \text{Attn}(q_{\text{str}}, k_{\text{str}}, v_{\text{str}})
\end{equation}
By restricting these queries to their respective subspaces, we establish a unidirectional information flow from the guidance sources to the target, ensuring that structural and reference attributes remain stable throughout the denoising process. Furthermore, this design significantly reduces computational overhead.

% For the target queries $q_{\text{tgt}}$, we use a frame-synced alignment path where query items are processed as frame-level representations $q_{\text{tgt},i}$. The attention search space for each frame $i$ is constrained to:
% \begin{equation}
% K_{\text{comp},i} = [K_{\text{tgt}}, k_{\text{str},i}, K_{\text{msk}}, K_{\text{ref}}]
% \end{equation}

% The identity reference branch $q_{\text{ref}}$ and mesh structure branch $q_{\text{str}}$ perform feature extraction exclusively within their own modalities, remaining isolated from the noisy target sequence:
% \begin{equation}
% x_{\text{ref}} = \text{Attn}(q_{\text{ref}}, k_{\text{ref}}, v_{\text{ref}}) \quad \text{and} \quad x_{\text{str}} = \text{Attn}(q_{\text{str}}, k_{\text{str}}, v_{\text{str}})
% \end{equation}
% This unidirectional information flow preserves the stability of reference features throughout the denoising process.

% This decomposition resolves the unconstrained competition between tokens that occurs in standard attention mechanisms.
\section{Experiment}

\subsection{Experiment Settings}
\subsubsection{Implementation Details}
Our framework is built upon the 14B parameter version of Wan2.1, a powerful DiT-based video diffusion model.
We employ Wan-2.1-T2V-14B~\cite{wan2025} as our base model due to its superior performance. We fine-tune all self-attention layers to accommodate multi-modal input conditions. We resize videos to 480×832 and sample 41 frames at 15 fps. Fine-tuning is conducted on 8 NVIDIA H200 GPUs for $20\text{K}$ steps, utilizing a learning rate of $2 \times 10^{-5}$ and a batch size of $8$. 
\subsubsection{Evaluation}
To facilitate a comprehensive quantitative evaluation, we introduce MozooBench, a novel benchmark. To ensure a fair comparison, none of data in the benchmark appear in our training data. We use the same construction method as the real-world dataset in the training data. 
We compared our work with the VACE~\cite{vace} and Recafade~\cite{huang2025refaccadeeditingobjectgiven} methods. Furthermore, considering that there is currently no controllable method for reference videos, we input reference images into the two methods mentioned above for comparison. To ensure a fair comparison, we further provide experimental results conditioned on image-based references.

To assess performance on this benchmark, we evaluate eight metrics across two aspects: video quality and reference alignment. Following prior work, we measure and assess video quality using subject and background consistency~\cite{huang2023vbench} to ensure identity preservation and environmental stability, motion smoothness~\cite{radford2021learningtransferablevisualmodels} to evaluate temporal coherence and detect jitter, and imaging quality~\cite{ke2021musiqmultiscaleimagequality} and aesthetic quality~\cite{schuhmann2022laion5bopenlargescaledataset} to provide a no-reference assessment of frame-level perceptual clarity and visual appeal. We also use widely used quantitative metrics to evaluate reference alignment, including PSNR and SSIM~\cite{wang2004image} to evaluate low-level structural similarity, along with LPIPS~\cite{zhang2018perceptual} to measure high-level perceptual distance between the generated output and the guidance sources.
\subsection{Quantitative Comparison}
We evaluate Mozoo against state-of-the-art methods using standard quantitative metrics. For a fair comparison, the VACE~\cite{vace} baseline is conditioned on the input video, a reference image, and a binary mask defining editable and fixed regions. As shown in Tab.~\ref{tab:comparison}, Mozoo consistently outperforms existing baselines in both structural fidelity and reference alignment. Whether using static images (I2V) or dynamic videos (V2V) as references, our framework achieves higher similarity to target sources while maintaining pixel-level accuracy. LPIPS scores further indicate that our approach produces textures with higher perceptual fidelity, narrowing the gap between coarse geometric proxies and photorealistic video.

\begin{table*}[t]
\centering
\caption{Quantitative results on MozooBench. \textbf{Bold} and \underline{underline} indicate the best and the second best performance, respectively.}
\label{tab:comparison}
\footnotesize
\setlength{\tabcolsep}{5pt} % 稍微调整列间距
\resizebox{\textwidth}{!}{
\begin{tabular}{l|ccccc|ccc} % 1 (Method) + 2 (Subject) + 3 (Video) + 2 (Human) = 8 列
\toprule
\multirow{2}{*}{\textbf{Method}}  & \multicolumn{5}{c|}{Video Quality} & \multicolumn{3}{c}{Reference Alignment} \\
\cmidrule(lr){2-6}  \cmidrule(lr){7-9}
& \makecell{ Subject \\ Consistency $\uparrow$} & \makecell{Background \\ Consistency $\uparrow$} & \makecell{Motion \\ Smoothness $\uparrow$} & \makecell{Imaging \\ Quanlity $\uparrow$} & \makecell{Aesthetic\\ Quality  $\uparrow$} & \makecell{PSNR$\uparrow$} & \makecell{LPIPS$\downarrow$} & \makecell{SSIM$\uparrow$}\\
\midrule
VACE & 93.68 & 96.28 & 99.15 & 69.34 & 55.82 & 15.628 & 0.984 & 0.901 \\
Refacade & 95.49 & 94.89 & \textbf{99.24} & 67.57 & 53.02 & 18.44 & 0.082 & 0.908 \\
\midrule
w/o RAR & 97.08 & 98.33 & 99.12 & 69.25 & 55.90 & 18.44 & 0.077 & 0.913\\
w/o ADA & 96.35 & 97.91 & 99.14 & 69.37 & 55.67 & 18.67 & 0.079 & 0.913\\
\midrule
\textbf{Ours (Image)} & \underline{97.52} & \underline{98.52} & 99.16 & \underline{69.39} & \underline{55.95} & \underline{19.61} & \underline{0.075} & \underline{0.915}  \\
\textbf{Ours (Video)} & \textbf{97.84} & \textbf{98.72} & \underline{99.20} & \textbf{69.41} & \textbf{56.30} & \textbf{20.75} & \textbf{0.070} & \textbf{0.922} \\
\bottomrule
\end{tabular}
}
\end{table*}

\subsection{Qualitative Comparison}
\begin{figure}[t] 
    \centering
    \includegraphics[width=\linewidth]{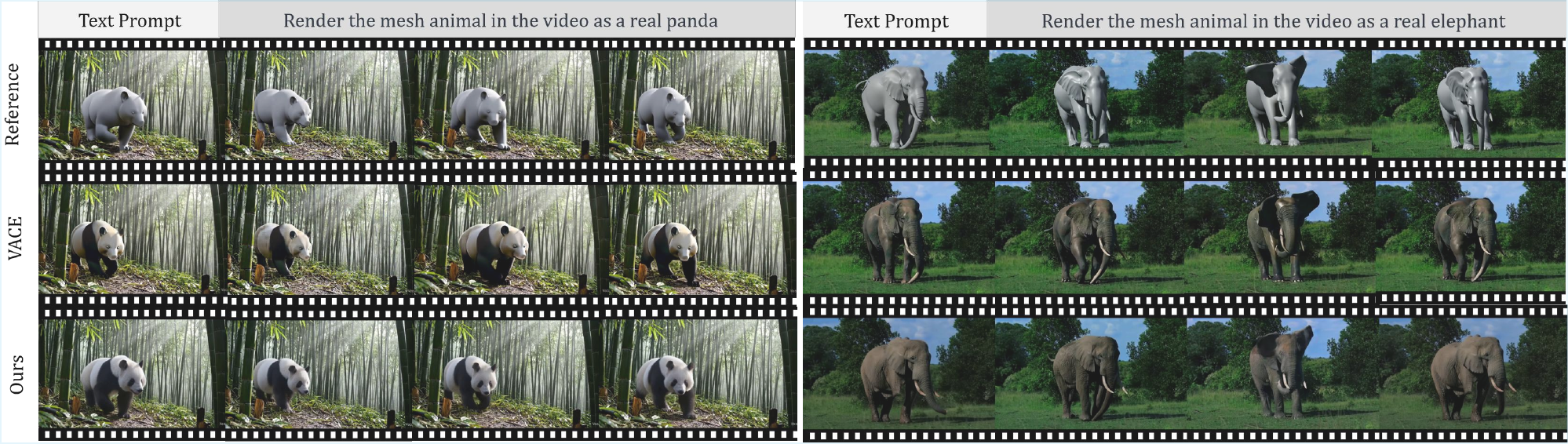}
    \caption{Qualitative comparison of text-based generation. Given text prompts and mesh-guided motion, our method synthesizes realistic animal textures with high fidelity. Notably, our approach preserves intricate high-frequency details, which appear over-smoothed in the VACE results.} % 修正了括号
    \label{fig:T2V_Comparison} % 确保标签唯一
\end{figure}

\begin{figure}[h] 
    \centering
    \includegraphics[width=\linewidth]{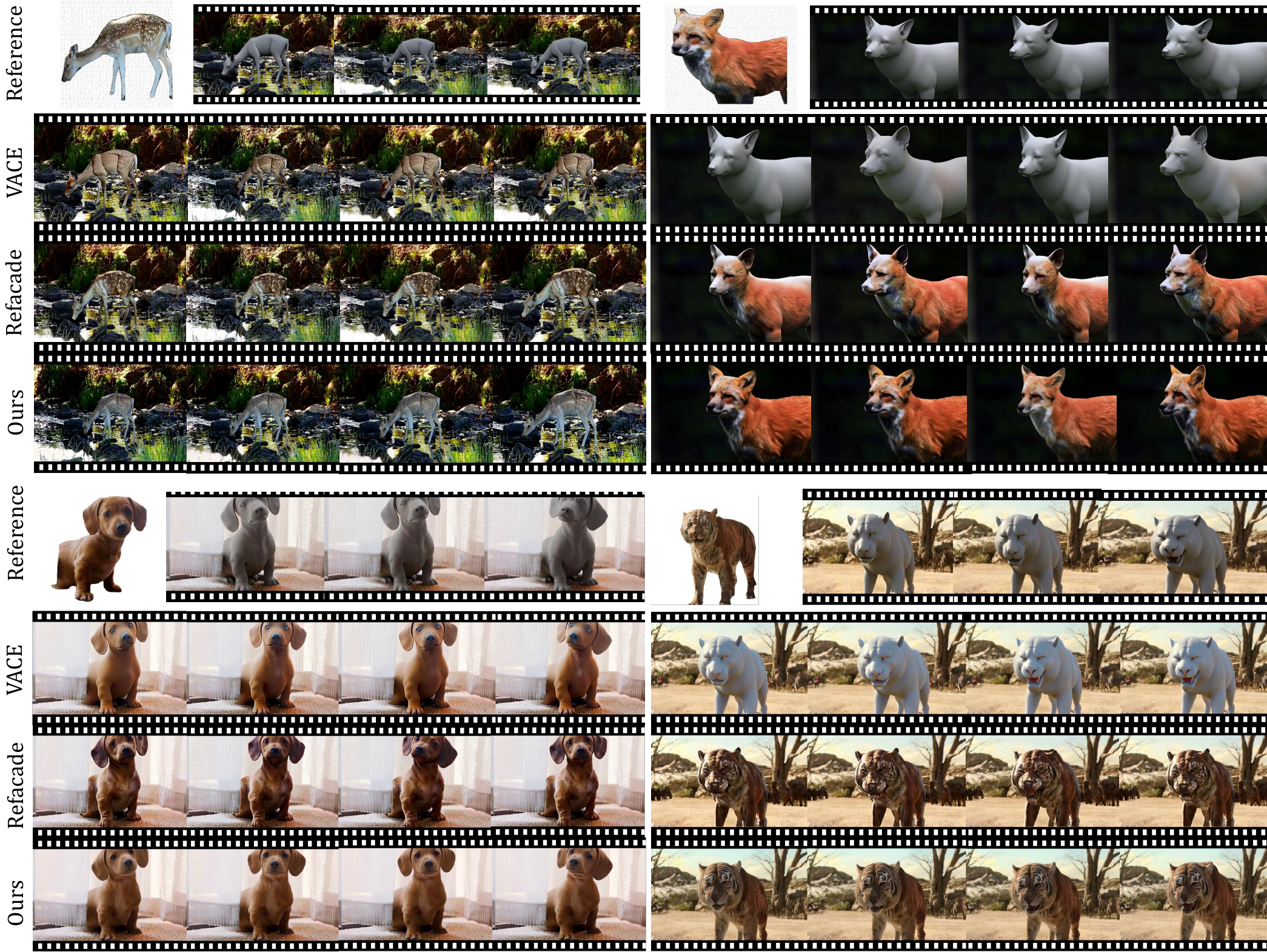}
    \caption{Qualitative comparison of reference-based generation. The results demonstrate that our proposed method can render animal subjects with superior spatio-temporal consistency, finer textural detail, and more realistic lighting effects.}
    \label{fig:I2V_Comparison} % 确保标签唯一
\end{figure}

As shown in Fig.~\ref{fig:T2V_Comparison} and Fig.~\ref{fig:I2V_Comparison}, we present two different feature comparisons for  text-to-video generation and  reference-to-video generation. Notably, while our model is capable of processing dynamic reference videos, we utilize a static reference image here to maintain a fair benchmark against prior methods. Our method shows robust versatility across different animal categories, delivering high-fidelity simulations of both muscle structures and complex fur textures. 

For text-based generation, VACE can transform animal subjects in mesh-guided videos into specific species, yet its results often appear flat when representing animals with fine fur. Our method improves textural depth and realism by capturing more intricate details. Similarly, in reference-based generation, while VACE and Refacade can simulate fur to some degree, both have clear limitations. For instance, VACE occasionally fails to synthesize a coherent subject, while Refacade often struggles to map textures accurately onto the correct anatomical regions. In contrast, our approach preserves fine fur details with higher visual fidelity.

\subsection{Ablation Study}
We conduct a comprehensive ablation study to evaluate our key architectural designs alongside the impact of different input modalities, specifically comparing image-conditioned (using a single reference image) and video-conditioned (using a reference video sequence) settings. To provide a clearer demonstration of the individual modules’ contributions, we employ the video-conditioned setting as the default configuration for the structural ablation analysis.
\subsubsection{Ablation on Architectural Components}
We analysis the individual contributions of our core designs: Role-Aware RoPE (RAR) and Asymmetric Decoupled Attention (ADA). As illustrated in Fig.~\ref{fig:ablation}, the absence of RAR leads to noticeable spatial drift in anatomical features; without the explicit coordinate mapping, the model fails to precisely anchor reference textures onto the target mesh. Furthermore, disabling ADA significantly compromises the reconstruction of high-frequency details. In the absence of the decoupled attention mechanism, subtle fur textures tend to be overwhelmed by dominant structural signals, resulting in blurred and over-smoothed visual outputs. These results underscore that the synergy between RAR and ADA is indispensable for achieving high-fidelity, structure-aware synthesis.

\subsubsection{Ablation on Reference Modalities}
\begin{figure}[t] 
    \centering
    \includegraphics[width=\linewidth]{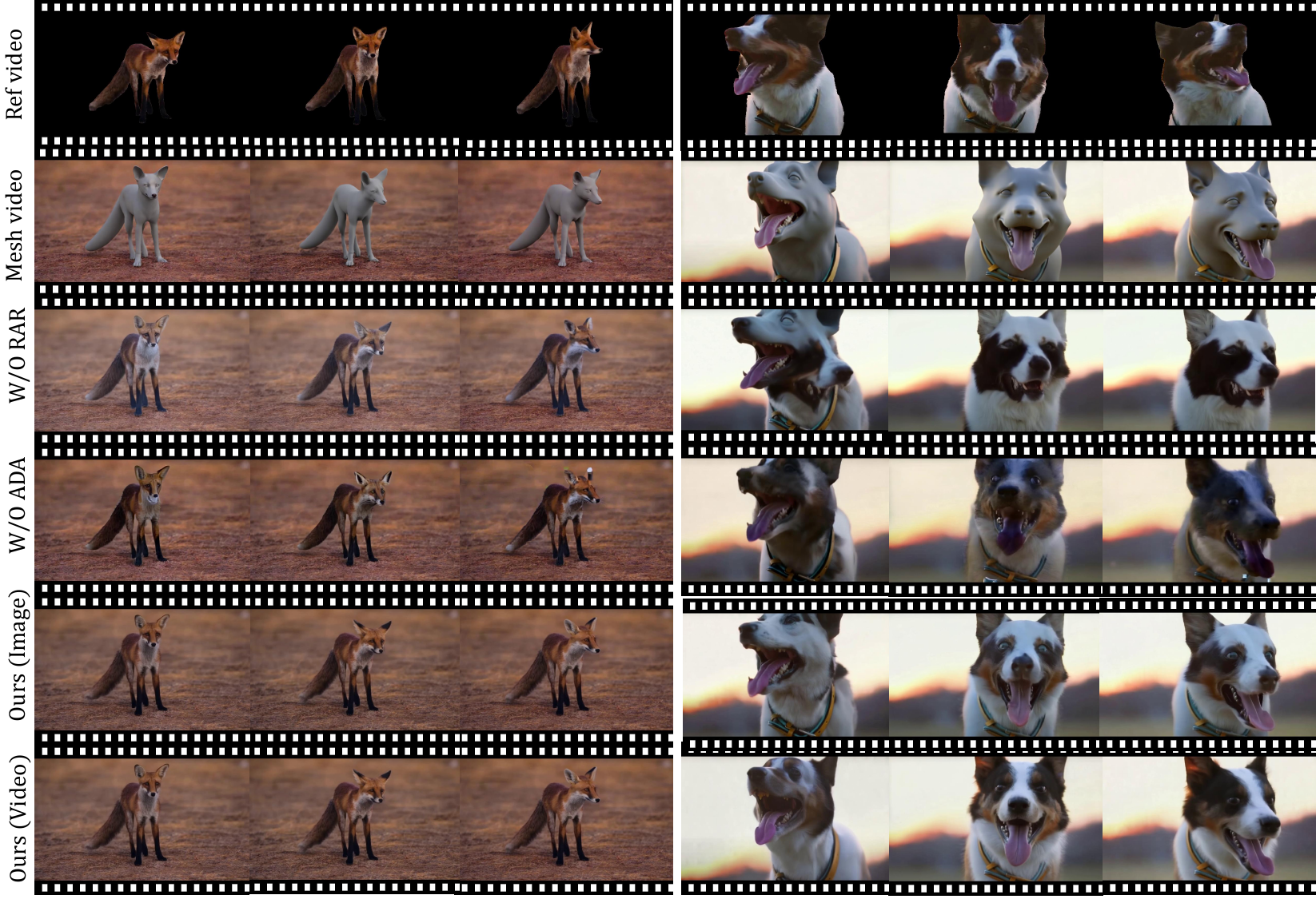}
    \caption{Ablation analysis of RAR and ADA in V2V Synthesis and comparison with I2V Modality. Removing these components leads to texture misalignment and loss of fine details. While our model achieves high quality results in both settings, the V2V configuration provides superior identity fidelity and texture consistency compared to the I2V setting.} % 修正了括号
    \label{fig:ablation} % 确保标签唯一
\end{figure}

Comparison between the two input configurations reveals that the video conditioned modality consistently outperforms the image conditioned setting across both video quality and reference alignment dimensions. While the image-to-video approach produces high-quality results, the video-to-video configuration achieves superior scores in terms of identity fidelity and temporal stability. As show in Fig.~\ref{fig:ablation}, while the performance of the  image conditioned model is comparable to that of the video conditioned configuration when the target pose aligns closely with the reference view, it exhibits limitations under significant viewpoint variations. When the generated motion covers extreme angles not present in the static reference image, the image condition approach may suffer from texture omissions or detail loss, whereas the video condition configuration effectively leverages temporal reference cues to maintain high-fidelity results across a wider range of perspectives.

\begin{figure}[h] 
    \centering
    \includegraphics[width=\linewidth]{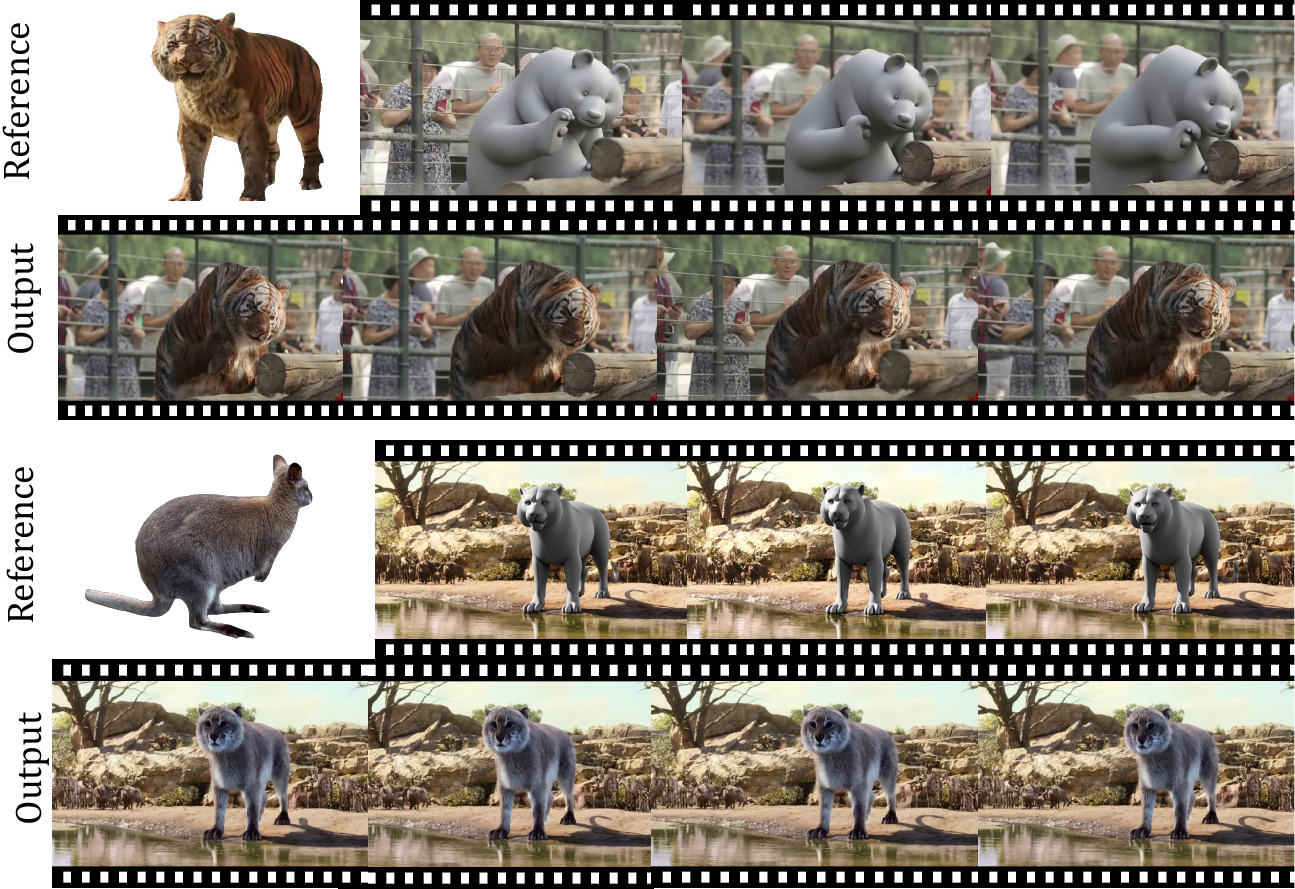}
    \caption{Results of cross-species texture transfer. Mozoo maps biological textures from a reference animal onto a source mesh proxy of a different species, achieving photorealistic synthesis with anatomical coherence.} % 修正了括号
    \label{fig:application} % 确保标签唯一
\end{figure}
\subsection{More Applications}
As shown in Fig.~\ref{fig:application}, our method can decouple high-level biological textures from their original anatomical structures and map them onto entirely different species. Mozoo ensures that the migrated surface attributes are biologically coherent, appearing as if they naturally originated from the target's own anatomy rather than being a superficial overlay. This zero-shot re-skinning paradigm facilitates the synthesis of high-fidelity, novel biological entities, providing an efficient and versatile tool for advanced character design and creative world-building in cinematic and interactive media.

\section{Conclusion}
In this paper, We present \textbf{MoZoo}, a generative framework that transforms coarse mesh sequences into photorealistic animal videos by integrating muscle and fur dynamics into an end-to-end process. By streamlining traditional refinement workflows, MoZoo significantly reduces manual effort in high-fidelity character production. We resolve multi-modal alignment and structural-textural interference via two key innovations: \textbf{Role-Aware RoPE (RAR-RoPE)}, which eliminates artificial temporal priors, and \textbf{Asymmetric Decoupled Attention}, which explicitly separates geometric guidance from textural details. Extensive evaluations on \textbf{MoZooBench} demonstrate MoZoo's superior textural fidelity and spatio-temporal consistency across diverse species.

Despite its efficacy, handling complex multi-animal interactions and severe occlusions remains a frontier. Future research will generalize MoZoo toward universal hair and fur simulation—including human hair and synthetic fibers—aiming for granular control over material attributes across diverse character archetypes beyond animal subjects. As the technology generalizes to human subjects, there is a risk of generating unauthorized digital avatars or deepfake videos, which may contribute to the spread of misinformation and undermine digital trust.

\clearpage  % TODO FINAL: This \clearpage needs to be removed from both review and camera-ready versions.

\bibliographystyle{splncs04}
\bibliography{main}
\end{document}